# Occluded Face Recognition Using Low-rank Regression with Generalized Gradient Direction

Cho-Ying Wu and Jian-Jiun Ding, *Senior Member, IEEE*

*Abstract*—In this paper, a very effective method to solve the contiguous face occlusion recognition problem is proposed. It utilizes the robust image gradient direction features together with a variety of mapping functions and adopts a hierarchical sparse and low-rank regression model. This model unites the sparse representation in dictionary learning and the low-rank representation on the error term that is usually messy in the gradient domain. We call it the "weak low-rankness" optimization problem, which can be efficiently solved by the framework of Alternating Direction Method of Multipliers (ADMM). The optimum of the error term has a similar weak low-rank structure as the reference error map and the recognition performance can be enhanced by leaps and bounds using weak low-rankness optimization. Extensive experiments are conducted on real-world disguise / occlusion data and synthesized contiguous occlusion data. These experiments show that the proposed gradient direction-based hierarchical adaptive sparse and low-rank (GD-HASLR) algorithm has the best performance compared to state-of-the-art methods, including popular convolutional neural network-based methods.

## I. INTRODUCTION

OCCLUSION is a real-world problem that would depreciate the performance of the face recognition [1]. To tackle with the occlusion situation, it is substantially important how we analyze image contents. Jia and Martinez [2] used the support vector machine to seek a hyperplane that is parallel to the affine space of the occluded data. Zhou *et al.* [3] introduced the Markov random field into the computation of sparse representation. Li *et al.* [4] proposed the sparse error coding (SSEC) to build a morphological graph model considering the structure of the error support. Liang and Li [5] further built models with discriminative error, structural error, and occlusion support based on the SSEC. He *et al.* [6] proposed the half-quadratic minimization framework and recovered the occluded data iteratively. Recently, sparse representation-based classification (SRC) [7] has been proposed to represent the data by the sparse linear combination of a dataset dictionary. Based on SRC, the collaborative representation-based classification

This work was supported Ministry of Science and Technology, ROC, under the contract of 103-2221-E-002 -121 -MY3.
C. Y. Wu and J. J. Ding are with Graduate Institute of Communication Engineering, National Taiwan University, 10617, Taiwan. (e-mail: r04942049@ntu.edu.tw, jjding@ntu.edu.tw).

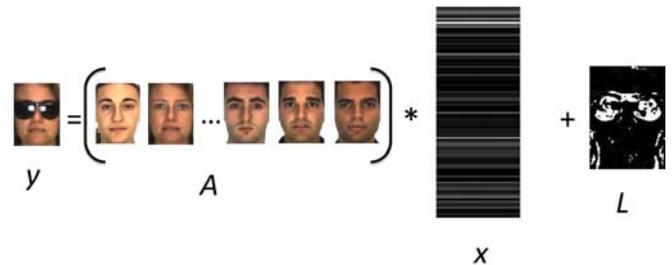

Fig. 1. The occlusion image is represented by a linear combination of all training data in the dictionary and added by a residual image standing for occlusion. *L* is the residue.

(CRC) [8] and the probability CRC (proCRC) [9] were proposed, which are helpful classification. Moreover, the robust sparse coding-based methods (RSC) [10], regularized robust coding (RRC) [11], and the fast version of RRC (F-IRNNLS) [12] solved the iteratively reweighted sparse coding problem and have effective results.

Lately, visual data with low-rank structure was extensively studied [13, 14]. The rank minimization problem is broadly utilized in matrix completion and the subspace clustering problem [13-16]. It was proven that the nuclear norm could be an effective convex surrogate of the rank terms in models [17]. Some methods use the rank minimization problem considering the low-rankness of the dictionary [18, 19] or the low-rankness of the occlusion part [20] to solve the occluded face recognition problem.

We find that the gradient directions with appropriate mapping functions is a more representative feature than the conventional image intensity domain operation [21] when using the sparse and low-rank model corroboratively.

In this paper, we propose the Gradient Direction-based Hierarchical Adaptive Sparse and Low-Rank (GD-HASLR) model. We first transform the operating domain from image intensity domain into the generalized image gradient direction domain by different S-shape mapping functions. We apply the horizontal and vertical filters up to three times and compute their gradient directions respectively. Next, we combine the sparsity constraint with the low-rank model, resulting to the sparse and low-rank regression model. Furthermore, we use the hierarchical adaptive weight on the sparsity constraint and show the relation between the proposed hierarchical adaptive sparse and low-rank model and the hierarchical adaptive lasso (HAL) [22, 23] model. Last, at the classification step, we design an effective poll strategy of all the gradient directions to further enhance the accuracy rate.

Experiments on real-world occlusion or disguise and synthesized occlusion data are included. We compare the results to other state-of-the-art methods. The proposed GD-HASLR algorithm has the best performance and is robust to synthesized block occlusions up to near 80% occluded face.

In the very recent researches, the convolutional neural network-based models (CNNs) [24-28] have been proposed to face recognition problems. We will also compare our method with the VGGFace net [24] and the lightened CNN [27, 28], which both have very high performance on occluded face recognition.

This paper is organized as follows: In section II, we will describe the backgrounds of the sparse representation and low-rank representation framework. In Section III, the proposed GD-HASLR method, including how to optimize the model, will be illustrated in detail. Section IV, extensive experiments are given and state-of-the-art methods are compared to validate the effect of the proposed algorithm. Finally, a conclusion is made in Section V.

## II. Preliminary

In this section, the background knowledge, including the gradient image, the sparse representation part, and low-rank representation, are explained. They are closely associated to the proposed GD-HASLR model introduced in the next section.

### A. Gradient Image

The gradient image [21] is usually used to detect the edges or features in an image. The gradient vector, the gradient magnitude, and the gradient direction are defined as follows:

$$\nabla f = [g_r, g_c]^T = \left[\frac{\partial f}{\partial r}, \frac{\partial f}{\partial c}\right]^T, \quad (1)$$

$$Magnitude(\nabla f) = \sqrt{g_r^2 + g_c^2}, \quad (2)$$

$$Direction(\nabla f) = \tan^{-1}(g_c / g_r) \quad (3)$$

where $r$ and $c$ are row and column directions, respectively. The gradient image is usually used as an effective feature of images applied to image content analysis and understanding [29-31].

### B. Sparse Representation

Suppose that we have a training face image dictionary set $A \in R^{d \times n}$ where $n$ is the number of samples from the training dataset and $d$ is the product of the image width and the height. Usually, training images are neutral faces of each identity without occlusion. Also, we have input samples $y \in R^d$ that are used as testing data. We can take a linear combination of all training samples to approximate the testing data with an error term $L \in R^d$:

$$y = Ax + L \quad (4)$$

where $x$ is the coefficient vector with dimension $n$. It is also plotted in Fig. 1.

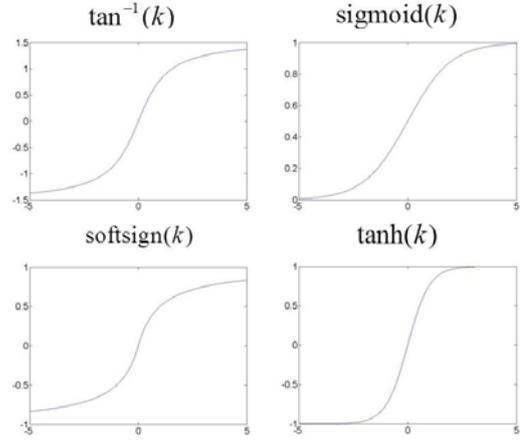

**Fig. 2.** Four kinds of S-shape functions are adopted as mapping functions where $k$ is an input variable.

Moreover, sparse representation adds the sparsity constraint to (4). From [7], face images within the same class would lie in the same feature subspace such that the occluded face data would be linearly correlated with the training images in the same class. This characteristic would make the solution $x$ sparse. Therefore, the problem in (4) can be reformulated as:

$$\min \|x\|_1 \quad s.t. \|y - Ax\|_2 \leq \varepsilon \quad (5)$$

where $\varepsilon$ is an error threshold. Using the Lagrangian method, (5) can be rewritten as an unconstrained optimization problem and solved by various methods [32]. After getting the sparse vector $x$, we calculate the $i^{th}$ class residue by

$$r_i = \|y - A\delta_i(x)\|_2 \quad (6)$$

where $\delta_i()$ is the $i^{th}$ class selector. Finally, we choose the class with the minimal residue.

### C. Low-rank Representation

The low-rank minimization problem is recently used in data processing and face recognition problem formulation. Some models apply the intrinsic low-rankness characteristic of data and decompose the corrupted data into the low-rank part and the occlusion part to construct a low-rank structure [18, 33][32]. Robust principal component analysis (RPCA) [33] tries to decompose the observation matrix $O$ into a low-rank matrix part $M$ and the corrupted error $E$ from the model as:

$$\min_{M,E} rank(O) + \lambda \|E\|_1 \quad s.t. \; O = M + E. \quad (7)$$

Low-rank representation (LRR) [13] is another model as follows:

$$\min_{Z,E} rank(Z) + \lambda \|E\|_1 \quad s.t. \; O = AZ + E \quad (8)$$

where $A$ is a dictionary whose atoms span the data space. Other works [12, 20] used the low-rank constraint visually and add the constraint on error term $L$ in (4) for considering that the occlusion part would occupy a certain contiguous area but not the whole face image. Consequently, the error term can be treated as an occlusion map that bearing a low-rank structure.

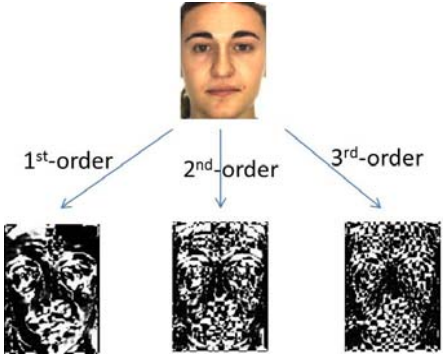

**Fig. 3.** A neutral face image from the AR database and its image gradient direction maps of the 1st, the 2nd, and the 3rd orders.

The low-rank regularized model in [20] is:

$$\min_{x,L} \|L\|_2^2 + \lambda \|L_M\|_* + \frac{1}{2}\eta x^T x \quad s.t. \quad y = Ax - L \quad (9)$$

where the fidelity term $\|L\|_2^2$ is the square of the $L_2$-norm, $\|L_M\|_*$ is the nuclear norm term that can be a surrogate of rank calculation [17], $L_M = Mat(L)$ is a conversion from a vector to a matrix with the original image size, the operator $M$ is the same for here and after, and $(1/2)\eta x^T x$ is a ridge regression term preventing the optimization from overfitting.

Optimization can be performed under the framework of the alternating direction method of multipliers (ADMM) [34] algorithm. Subsequently, one will get the coefficient vector $x$ and the occlusion map $L$ from (4) and the residual is:

$$r_i(y) = \|(y - A\delta_i(x) - L)_M\|_* \quad (10)$$

where $\delta_i(.)$ is the $i^{th}$ class selector. Finally, we select the class with the minimum residue as the face identity.

### III. PROPOSED GRADIENT DIRECTION-BASED ADAPTIVE SPARSE AND THE LOW-RANK MODEL

In this section, we explain the proposed model on how to model the atom coefficient vector $x$ with the training dictionary $A$ and the test data $y$ in (4) and how to optimize the proposed model. Also, we will explain the output-deciding algorithm during classification.

#### A. Generalized Image Gradient Direction

The proposed method is performed on the generalized image gradient direction domain. Different from the conventional definition in (3), we determine the generalized image gradient direction whose mapping function from the image intensity domain to the gradient domain is not confined to the usual one, $\tan^{-1}(k)$ [21]. Functions with increasing S-shape which converge to a positive value when $k \to \infty$ and converge to a negative value when $k \to -\infty$ can also be mapping functions. The convergence is needed because when $g_c/g_r$ in (4) is unbounded, the ADMM optimization process may fail to converge. Some desirable mapping functions are proposed in Fig. 2, including $\tan^{-1}(k)$, $\tanh(k)$, the soft-sign function is $k/(1+|k|)$, and the sigmoid function is $1/(1+e^{-k})$. The input variable $k$ in these functions can be scaled and shifted. These functions are more flexible to attain high recognition performance.

Here, we use the sigmoid function for explanation. The mapping function with input image $I$ is as:

$$\phi(\nabla g(I), u, v) = \frac{1}{1 + e^{-u(\nabla g(I) - v)}} \quad (11)$$

where $\nabla g(I) = g_c(I)/g_r(I)$, $g_c = h_c * I$, $g_r = h_r * I$, $*$ is the convolution operator, $h_r$ and $h_c$ are horizontal and vertical filters to calculate the gradients along row and column directions, respectively (Here, 3x3 Sobel filters [21] are applied), and $u$ and $v$ are scaling and shifting parameters.

Based on the knowledge that gradients are effective features in image content analysis and understanding, we propose to use the following feature vector

$$f = \left[ \nabla g, \nabla^2 g, \nabla^3 g \right]^T$$

where $\nabla^2 g = \nabla(\nabla g)$ and $\nabla^3 g = \nabla(\nabla^2 g)$ are the second-order and third-order image gradients constructed from $\nabla g$ defined in (11). The first-order to the third-order gradients are applied to constitute the gradient vector $f$. Then, the gradient direction vector is

$$\Phi = \left[ \phi(\nabla g), \phi(\nabla^2 g), \phi(\nabla^3 g) \right]^T$$

where $\phi$ is defined in (11). Illustration of a face image mapped onto the generalized gradient direction domain by the sigmoid function is shown in Fig. 3.

#### B. Sparse and Low-Rank Regularized Regression

In (4), it is assumed that there is a dataset with $d$-dimension and $n$ samples that consist of a training dictionary $A \in R^{d \times n}$ and a testing data $y \in R^d$. One can represent $y$ by taking a linear combination of each atom in $A$ with the error term $L$.

From the gradient direction vector $\Phi$ containing the 1st, the 2nd, and the 3rd order gradient direction features, the test input image $y$ in (4) can be replaced with its gradient direction feature $\phi(\nabla g(y))$, $\phi(\nabla^2 g(y))$, and $\phi(\nabla^3 g(y))$ (for simplicity, we denote them as $y_{g1}$, $y_{g2}$, and $y_{g3}$, respectively. The similar notations will be applied to the dictionary $A$ and the error $L$). Thus, one will get three output $x$ vectors: $x_{g1}$, $x_{g2}$, and $x_{g3}$, in (4) for a single testing sample $y$ from solving the regression model introduced as follows.

In the sequel, $x_i$ is used for denoting the $i^{th}$ component of a general vector $x$ that can either be $x_{g1}$, $x_{g2}$, or $x_{g3}$. The similar notations will be applied to $y$, $A$, and $L$ that can be either imposed with $g_1$, $g_2$, or $g_3$.

As in [12, 20], the rank is a powerful metric to describe the structural of the error term $L \in R^d$. In (9), it is assumed that $L$ is low-rank since many rows and columns are zero in the occlusion map. The term of $(1/2)\eta x^T x$ avoids overfitting on the occluded face recognition problem. Eqs. (7) and (8) combined sparsity and low-rank constraints in the matrix completion problem. In [35], the sparsity constraint and the low-rank constraint are combined to solve the video emotion recognition problem.

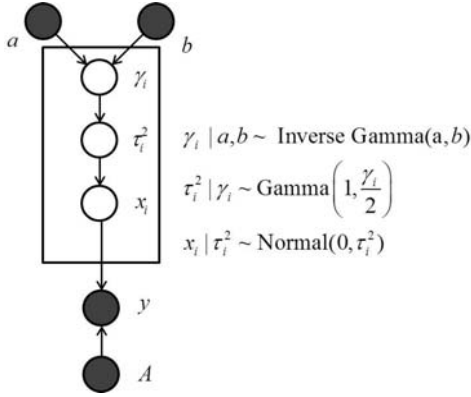

**Fig. 4.** Direct graphical model and the conditional distribution relation between latent nodes for HAL. Solid rounds are observed variables and hollow rounds are latent variables.

In this work, to solve the occluded face recognition problem, the collaborative constraints, which is a combination of the sparsity and low-rank constraints, is utilized to construct the following model:

$$\min_{x,L} \|x\|_1 + \alpha \|L_M\|_* \quad \text{s.t.} \quad y = Ax + L \qquad (12)$$

where $\alpha > 0$ is the rank penalty term and $L_M$ is the same as that in (9).

This is the regularization problem of (4) and the symbol here is consistent to Section II. To explain the regularization, considering that we desire the solution of $x$ to be discriminative, we impose the sparsity constraint on $x$. Moreover, we desire the occlusion map $L$ to have a non-zero value only in certain areas, as in Fig. 1. Therefore, the low-rank constraint on $L$ using the nuclear norm is imposed.

### C. Hierarchical Adaptive Sparse and Low-Rank Regression

Based on the sparse representation framework in (4) and (5) and consistent symbols $x$, $y$, and $A$, in the proposed algorithm, the HAL prior uses the graphical model of three latent variables in a hierarchical way: the mean of the Gamma distribution ($\gamma_i / 2$), the variance of the mixture of Gaussian distributions $\tau_i^2$, and the coefficient $x_i$ corresponding to the dictionary $A$. Their relations are shown in Fig. 4. After integrating $\tau_i^2$, one can get $x_i$, which follows the generalized $t$-distribution where $i$ denotes the $i^{\text{th}}$ component:

$$p(x_i \mid a, b) = \frac{a}{2b} \exp\left(\frac{|x_i|}{b} + 1\right)^{-(a+1)}. \qquad (13)$$

Applying maximum a posteriori (MAP) to the prior in (13) is an $L_1$-penalized optimization problem with logarithmic penalization. We compute MAP estimation using the expectation maximization (EM) algorithm and the $M$-step is a weighted lasso problem:

$$\hat{x}^{t+1} = \arg\min_x \|y - Ax^t\|_2^2 + \sum_i s_i^t |x_i^t| \qquad (14)$$

where $s_i^t = E[1/\gamma_i] = (a+1)/(b+|x_i^t|)$ is calculated in the E-step for the prior in (13), and $t$ denotes the iteration number in the EM optimization process.

The hierarchical adaptive lasso model can utilize different $p(x_i)$ and applies the following objective function:

$$\hat{x}^{t+1} = \arg\min_x \|y - Ax^t\|_2^2 + \sum_i \pi_\lambda(|x_i^t|) \qquad (15)$$

where $\pi_\lambda(|x_i|) = -\log p(x_i \mid \lambda)$. Here, $\lambda = (a, b)$ is the tuning parameter that makes (15) be of logarithmic penalization. Then, by combining the HAL model and the sparse and low-rank regression model in (12), the following objective function is obtained:

$$\min_{x,L} \alpha \|L_M\|_* + \sum_i \pi_\lambda(|x_i|) \quad \text{s.t.} \quad y = Ax + L. \qquad (16)$$

We call it Hierarchical Sparse and Low-Rank Regression. It has the advantages of retaining low-rankness and obtaining adaptive weight penalty on the sparse penalty term and has a better solution for $x$ to further enhance the performance. We will show the relation of (16) and the HAL model in (14) later.

### D. Optimization

Then, we solve (10) to obtain $x$ and $L$ by the alternating direction method of multipliers (ADMM) [16]. Following the framework of the ADMM, the augmented Lagrangian function of (16) can be derived as:

$$L(x, L, z) = \alpha \|L_M\|_* + \sum_i \pi_\lambda(|x_i|) \\ + z^T(y - Ax - L) + \frac{\beta}{2} \|y - Ax - L\|_2^2 \qquad (17)$$

where $z$ is the Lagrange multiplier and $\beta > 0$ is the penalty parameter.

With some simplification of the inner product and the $L_2$-norm square, the updating at the $k^{\text{th}}$ iteration of ADMM can be expressed by the following three sub-problems:

$$L^{k+1} = \arg\min_L \alpha \|L_M\|_* + \frac{\beta}{2} \|y - Ax^k - L + \frac{1}{\beta} z^k\|_2^2, \qquad (18)$$

$$x^{k+1} = \arg\min_x \sum_j \pi_\lambda(|x_j|) + \frac{\beta}{2} \|y - Ax - L^{k+1} + \frac{1}{\beta} z^k\|_2^2, \qquad (19)$$

$$z^{k+1} = z^k + \beta(y - Ax^{k+1} - L^{k+1}). \qquad (20)$$

From [36], singular value thresholding (SVT) is an effective method to solve the nuclear norm minimization problem. Therefore, (18) can be solved by

$$(U, S, V) = SVD(y - Ax^k + \frac{1}{\beta} z^k),$$

$$L^{k+1} = U\left(\text{soft}(S, \frac{\alpha}{\beta})\right) V^T \qquad (21)$$

where $\text{soft}(.,.)$ is the soft thresholding operator [37]. Eq. (19) has the form the same as (15). Therefore, the hierarchical sparse and low-rank regression reduces to HAL through ADMM sub-problem formulation via the augmented Lagrangian method. Eq. (19) can be optimized by the existing fast lasso solver [17] with the adaptive penalty term.

The ADMM iteration process is set to be convergent when $(\|x^{k+1}\|_2 - \|x^k\|_2)/\|x^k\|_2 < 10^{-6}$ and $(\|L^{k+1}\|_2 - \|L^k\|_2)/\|L^k\|_2 < 10^{-6}$ are both fulfilled.

### E. Residual Computation and Classification

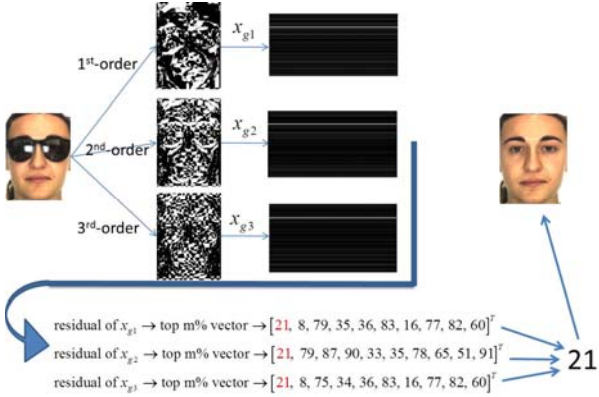

**Fig. 5.** A sunglasses occluded face in the AR database. In sum, three orders of gradient directions are computed. Ideal sparse *x* vectors are as shown. The most frequent class number appearing in every top 10% vectors is 21. It has frequency = 3 and is selected as the identity.

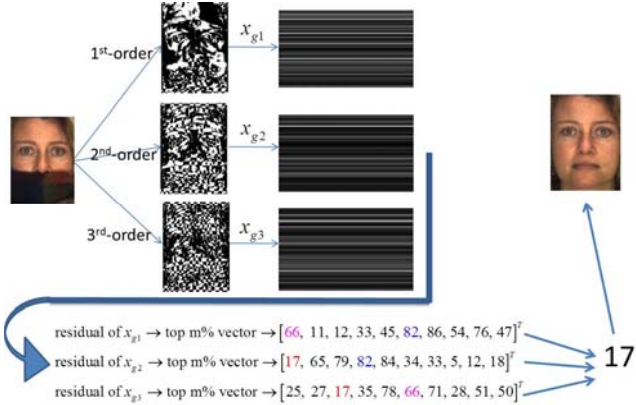

**Fig. 6.** A scarf occluded face in the AR database. Different from Fig. 3, this shows how to vote and make the decision. In this case, there are three most frequent class numbers with frequency = 2. However, the class number 17 has the least average rank and is selected as the identity.

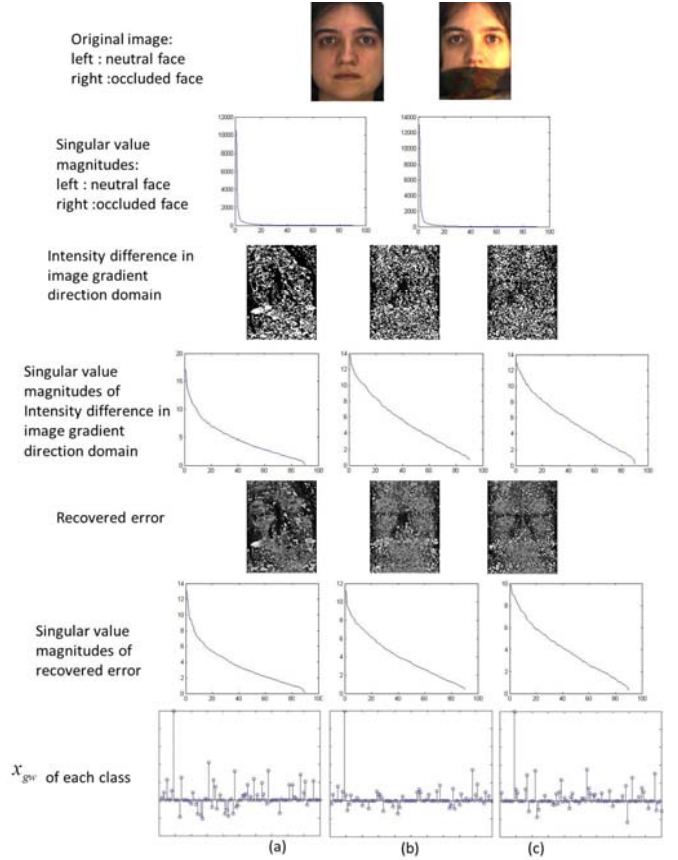

**Fig. 7.** Low-rankness analysis in the image gradient direction domain. Three gradient direction features are adopted. (a) Using the first-order feature. (b) Using the second-order feature. (c) Using the third-order feature. The singular values in the second, the fourth, and the sixth rows are calculated from singular value decomposition (SVD).

The obtained $x_{g1}$, $x_{g2}$, and $x_{g3}$ specify the solutions for the first-order, the second-order, and the third-order gradient direction features for a single test sample $y$ from

$$\hat{x}_{gw} = \arg\min_x \| y_{gw} - A_{gw}x_{gw} \|_2^2 + \sum_i \pi_\lambda(|x_{gw,i}|), \quad w=1,2,3. \quad (22)$$

The residual is computed from:
$$r_{n,gw}(y_{gw}) = \| (y_{gw} - A_{gw}\delta_n(\hat{x}_{gw}) - L_{gw})_M \|_*, \quad w=1,2,3 \quad (23)$$

where $n$ is in the range of all the class numbers and $\delta_n()$ is a selector to identify the class $n$. Then, the proposed classification method as follows is applied. First, the residues are rearranged in an ascending way, that is, from the least to the largest residue. Then, the top $m$% residues are selected.

Note that the correct class label may not be always at the first or the second place with the minimal residue, but usually appears in the top $m$% vectors and has the highest appearing frequency. In this sense, the outlier occupying the first place in one residual vector but ranking behind in the other two residual vectors will not be considered to be a correct label. Hence, we calculate the number of occurrences for each class in the three top $m$% residual vectors and choose the most frequent one. If two or more class numbers have the same occurring frequency, we select the one with the least average rank. Fig. 5 and Fig. 6 are two examples to explain the voting strategy.

### F. Low-rankness in Image Gradient Direction Domain

In this section, we explain the low-rankness assumption of the error term $L$ in the image gradient direction domain. In Fig. 1, in the image intensity domain, the low-rankness assumption is reasonable since the error term reveals a spatially contiguous occlusion object. However, in the image gradient direction domain, the difference map (shown in the third row of Fig. 7) is non-contiguous and has non-trivial values scattered in the map. Thus, the spatial low-rankness assumption may not hold. This can be seen from the 4th row in Fig. 7 that the singular value magnitude curve of the intensity difference image may not drop down as sharply as a natural image, as the 2nd row in Fig. 7. We call this phenomenon "weak low-rankness". However, the model in (16) is still applicable to the error term $L$ with weak low-rankness and does not depreciate the effectiveness of minimum nuclear norm optimization. Soft thresholding in (22), updates $Z_k$ in (20) and singular value decomposition of $y - Ax_k + (1/\beta)Z_k$ in (21) will collaboratively make the error $L$ converge to the rankness close to the intensity difference map (see the 4th row and the 6th row in Fig. 7). Also, the $L_1$ penalty for $x$-updating in (19) retains the sparsity of $x$.

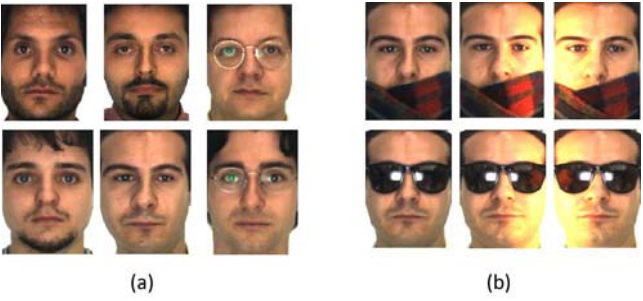

**Fig. 8.** Samples from the AR database. (a) The neutral image set in session 1 for training. (b) The 6 occluded face images of a person with different light conditions in session 2 for testing.

In Fig. 7, observing the 3$^{rd}$ row and the 5$^{th}$ row, one can see that the recovered error has similar spatial structure with the intensity difference map. The blurred pixel in the 5$^{th}$ row is the footprint by soft thresholding. It reveals that, in the weak low-rankness optimization problem, the recovered structure will not be perfectly similar to the ideal one compared to the application of singular value thresholding to the matrix completion problem [36].

However, we aim at recognizing the identity of occluded faces. By losing the information of the occlusion object in the image gradient direction domain, the recognition rate can be further enhanced. Therefore, one may say that the weak low-rankness property has advantages in the recognition problem since the effect of the occlusion object is reduced. Last, in the 7$^{th}$ row, we reveal the sparsity of $x$.

The overall procedure of the proposed GD-HASLR algorithm is described in Algorithm 1. Its performance will be validated by the simulations in the next section.

---

**Algorithm 1:** Proposed Gradient Direction-based Hierarchical Sparse and Low-Rank Model (GD-HASLR)

**Input**: training dictionary $A$, testing image $y$
1) For $A$ and $y$, calculate their generalized gradient directions of three orders, denoted by $A_{g1}, A_{g2}, A_{g3}$ and $y_{g1}, y_{g2}, y_{g3}$.
2) **For** $w = 1$ to 3:
3)   Solve (17) with training-testing pairs ($A_{gw}, y_{gw}$). (Initially, set $x = 0$, $L = 0$, $z = 1$, and $k = 0$).
4)   **While** not converge:
5)     $k = k+1$
6)     Update $L_{gw}^{k+1}$ by (18)
7)     Update $x_{gw}^{k+1}$ by (19), using the adaptive weight the same as that in (15)
8)     Update $z^{k+1}$ by (20)
9)   **Until** converge
10)  Calculate residues $r_{n,gw}$ by (23) for all class number $n$.
11) **End For**
12) Extract top $m$% residues from $r_{g1}, r_{g2}$, and $r_{g3}$.
13) Poll the identity from residues with the most frequently occurring index. When two or more index numbers have the same frequency, select the index number with the least average rank.

**Output**: The final decided identity number

---

## IV. Experiments on Real Disguise and Block Occlusion

In this section, extensive experiments are conducted on occlusion face recognition.

About parameter setting, the tangent hyperbolic function is chosen as the mapping function to compute the generalized image gradient direction. In (11) and (17), $u = 7.3$, $v = 0.51$, $\alpha = 100$, and $\beta = 1$. The top 10% index numbers are extracted during classification. The Normal Inverse Gaussian (NIG) [31, 38] function as follows is used as the adaptive weight function for the hierarchical model:

$$\pi(x_i) = \log\left(\sqrt{\delta^2 + x_i^2}\right) - \log\left(K_1\left(\gamma\sqrt{\delta^2 + x_i^2}\right)\right) \quad (24)$$

where $K_1$ is the 1$^{st}$ order modified Bessel function of the second kind. Notice that (24) is a simplification form of the NIG function. Here, we set $\delta = 1$ and $\gamma = 10^{-6}$.

### A. AR Database Real-world Occlusion

First, we use the occluded face data (sunglasses and scarf from session 1 and session 2) in the AR database [39]. The AR database is a complete and famous database focusing on occluded face data. It consists of 100 individuals and two sessions with total 26 images for each person. For each individual, 1 neutral face image, 3 sunglasses occluded faces, and 3 scarf occluded faces are treated as the testing data. Each sunglasses or scarf occluded face image suffered from different kinds of illumination conditions. Moreover, the synthesized contiguous occlusion data are used, representing the irregular contiguous occlusion appearing in any possible location on the face. We compare the results with other state-of-the-art methods. All images are resized to 42×30 pixels. All algorithms were implemented in Matlab. We use only one neutral face image of each person from session 1 as the training face. Different sets of occluded face images from sessions 1 and 2 (total 6 images for each person each session test) are used for testing. Some samples are shown in Fig. 8.

Simulation results are shown in Table I. We compare the proposed algorithm to state-of-the-art methods as NR[20], WeiTIP [19], HQPAMI [6], proCRC [9], RRC [11], RSC [10], SSEC [4], and F-LR-IRNNLS [12]. We have all tuned the parameters of these methods to the optimal case.

From Table I, one can observe that, in overall, the proposed GD-HASLR algorithm has the highest recognition rate. It outperforms other methods for **at least 8.58%** of the accuracy rate. For each testing subset, the proposed algorithm all achieve the highest recognition rate. It proves that the proposed GD-HASLR algorithm has high robustness against real-world disguise.

This experiment is challenging because it has unbalanced training data versus testing data (one versus six each person), and containing no occlusion information in the training dataset. However, this is maybe the most common scenario in the real-world that we have only one image of a person coping with any possible camouflage of the targets.

Second, we use two neutral images of each person from both sessions 1 and 2. We use the same testing sets as those of the first experiment. The results are shown in Table II.

TABLE I

ACCURACY ON THE AR DATABASE WHEN ONLY ONE NEUTRAL FACE IMAGE (FROM THE FIRST SESSION) OF EACH PERSON IS USED AS THE TRAINING DATA AND DIFFERENT OCCLUSION SETS ARE USED AS THE TESTING DATA.

| Methods | Sunglasses | | Scarf | | Overall |
|---|---|---|---|---|---|
| | Session1 | Session2 | Session1 | Session2 | |
| NR [10] | 0.2833 | 0.1667 | 0.2967 | 0.1733 | 0.2300 |
| WeiTIP [19] | 0.2067 | 0.1333 | 0.0667 | 0.0567 | 0.1159 |
| HQPAMI [5] | 0.5667 | 0.3800 | 0.3800 | 0.2233 | 0.3875 |
| ProCRC [3] | 0.5307 | 0.3100 | 0.1867 | 0.0733 | 0.2752 |
| RRC_$L_1$ [20] | 0.7500 | 0.5433 | 0.6033 | 0.4033 | 0.5750 |
| RRC_$L_2$ [20] | 0.7667 | 0.5367 | 0.5933 | 0.3867 | 0.5709 |
| RSC [4] | 0.6767 | 0.5133 | 0.6833 | 0.4967 | 0.5925 |
| SSEC [6] | 0.7067 | 0.4167 | 0.7033 | 0.5700 | 0.5992 |
| F-LR-IRNNLS [22] | 0.8867 | 0.6033 | 0.6700 | 0.4967 | 0.6642 |
| Proposed GD-HASLR | **0.9200** | **0.6667** | **0.8267** | **0.5867** | **0.7500** |

TABLE II

ACCURACY ON THE AR DATABASE WHEN TWO NEUTRAL FACES FROM THE TWO SESSIONS OF EACH PERSON ARE USED AS THE TRAINING DATA AND DIFFERENT OCCLUSION SETS ARE USED AS THE TESTING DATA.

| Methods | Sunglasses | | Scarf | | Overall |
|---|---|---|---|---|---|
| | Session1 | Session2 | Session1 | Session2 | |
| NR [10] | 0.3400 | 0.3333 | 0.3300 | 0.3567 | 0.3400 |
| HQPAMI [5] | 0.6133 | 0.5933 | 0.4467 | 0.4800 | 0.5333 |
| ProCRC [3] | 0.5300 | 0.5467 | 0.1800 | 0.1767 | 0.3584 |
| RRC_$L_1$ [20] | 0.8200 | 0.8433 | 0.6800 | 0.6533 | 0.7492 |
| RRC_$L_2$ [20] | 0.8100 | 0.8533 | 0.6867 | 0.6400 | 0.7475 |
| RSC [4] | 0.8000 | 0.8467 | 0.7633 | 0.7633 | 0.7933 |
| SSEC [6] | 0.7500 | 0.7367 | 0.7667 | 0.7500 | 0.7509 |
| F-LR-IRNNLS [22] | 0.9033 | 0.8767 | 0.7867 | 0.7600 | 0.8317 |
| Proposed GD-HASLR | **0.9300** | **0.9333** | **0.8267** | **0.8400** | **0.8825** |

TABLE III

ACCURACY OF THE PROPOSED GD-HASLR ALGORITHM AND THE CNN-BASED METHODS. ONE OR TWO NEUTRAL FACES OF EACH PERSON ARE USED AS THE TRAINING DATA AND DIFFERENT OCCLUSION SETS OF EACH PERSON ARE USED AS THE TESTING DATA. FC6 AND FC7 STAND FOR DIFFERENT FEATURES EXTRACTED FROM DIFFERENT FULLY-CONNECTED LAYERS. A AND B STAND FOR MODELS A AND B. FEATURES ARE CLASSIFIED USING NEAREST NEIGHBORHOOD METHODS.

| Methods | Sunglasses | | Scarf | | Overall |
|---|---|---|---|---|---|
| | Session1 | Session2 | Session1 | Session2 | |
| Using one neutral face from session 1 as training data of each person. | | | | | |
| VGG-Face FC6 [24] | 0.5400 | 0.4500 | **0.9167** | **0.8800** | 0.6967 |
| VGG-Face FC7 [24] | 0.4567 | 0.4000 | 0.8867 | 0.8400 | 0.6459 |
| Lightened CNN (A) [41] | 0.6733 | 0.5600 | 0.8700 | 0.8233 | 0.7317 |
| Lightened CNN (B) [42] | 0.3633 | 0.3133 | 0.8067 | 0.7367 | 0.5550 |
| Proposed GD-HASLR | **0.9200** | **0.6667** | 0.8267 | 0.5867 | **0.7500** |
| Using two neutral faces from session 1 and session 2 as training data of each person. | | | | | |
| VGG-Face FC6 [24] | 0.4467 | 0.5100 | **0.9167** | **0.9333** | 0.7017 |
| VGG-Face FC7 [24] | 0.4167 | 0.4467 | 0.8867 | 0.8933 | 0.6608 |
| Lightened CNN (A) [41] | 0.6467 | 0.5833 | 0.8667 | 0.8533 | 0.7375 |
| Lightened CNN (B) [42] | 0.3867 | 0.3800 | 0.8167 | 0.7933 | 0.5942 |
| Proposed GD-HASLR | **0.9300** | **0.9333** | 0.8267 | 0.8400 | **0.8825** |

TABLE IV
ACCURACY USING DIFFERENT ORDER IMAGE GRADIENT DIRECTIONS AS FEATURES.

| Methods | Sunglasses | | Scarf | | Overall |
|---|---|---|---|---|---|
| | Session1 | Session2 | Session1 | Session2 | |
| Using one neutral face from session 1 as training data of each person. | | | | | |
| Intensity domain | 0.2733 | 0.3867 | 0.1600 | 0.1800 | 0.2500 |
| GD-HASLR 1st-order | 0.7500 | 0.5000 | 0.6533 | 0.4233 | 0.5817 |
| GD-HASLR 2nd-order | 0.8600 | 0.6067 | 0.7467 | 0.4233 | 0.6592 |
| GD-HASLR 3rd-order | 0.7967 | 0.5100 | 0.6800 | 0.4300 | 0.6042 |
| Proposed GD-HASLR | **0.9200** | **0.6667** | **0.8267** | **0.5867** | **0.7500** |
| Using two neutral faces from session 1 and session 2 as training data of each person. | | | | | |
| Intensity domain | 0.3100 | 0.3433 | 0.3367 | 0.3300 | 0.3300 |
| GD-HASLR 1st-order | 0.7633 | 0.7800 | 0.6600 | 0.6500 | 0.7133 |
| GD-HASLR 2nd-order | 0.8433 | 0.8800 | 0.7600 | 0.7233 | 0.8017 |
| GD-HASLR 3rd-order | 0.8200 | 0.8133 | 0.7233 | 0.6733 | 0.7575 |
| GD-HASLR Polling | **0.9300** | **0.9333** | **0.8267** | **0.8400** | **0.8825** |

TABLE V
ACCURACY USING DIFFERENT MAPPING FUNCTIONS TO CALCULATE GENERALIZED IMAGE GRADIENT DIRECTIONS.

| Methods | Sunglasses | | Scarf | | Overall |
|---|---|---|---|---|---|
| | Session1 | Session2 | Session1 | Session2 | |
| Using one neutral face from session 1 as training data of each person. | | | | | |
| $\tan^{-1}(x)$ | 0.9100 | 0.6767 | 0.8067 | 0.6033 | 0.7492 |
| $\tanh(x)$ | 0.9200 | 0.6667 | 0.8267 | 0.5867 | 0.7500 |
| softsign(x) | 0.9033 | 0.6667 | 0.8100 | 0.6033 | 0.7458 |
| sigmoid | 0.9067 | 0.6700 | 0.7867 | 0.5633 | 0.7317 |
| Using two neutral faces from session 1 and session 2 as training data of each person. | | | | | |
| $\tan^{-1}(x)$ | 0.9067 | 0.9300 | 0.8367 | 0.8200 | 0.8733 |
| $\tanh(x)$ | 0.9300 | 0.9333 | 0.8267 | 0.8400 | 0.8825 |
| softsign(x) | 0.9067 | 0.9233 | 0.8467 | 0.8233 | 0.8750 |
| sigmoid | 0.9067 | 0.9100 | 0.8233 | 0.8100 | 0.8625 |

TABLE VI
ACCURACY WITH REGULARIZED WEIGHTS FROM DIFFERENT FEASIBLE FUNCTIONS (I.E. $\pi_\lambda$ IN (17)).

| Methods | Constant (set as 20) | Laplace | Generalized student T (HAL) | Normal Inverse Gaussian |
|---|---|---|---|---|
| Session 1 sunglasses | 0.8067 | 0.8133 | 0.8067 | 0.8276 |
| Session 1 scarf | 0.8867 | 0.8733 | 0.8900 | 0.9200 |
| Session 2 sunglasses | 0.5700 | 0.5767 | 0.5700 | 0.5867 |
| Session 2 scarf | 0.6367 | 0.6367 | 0.6400 | 0.6667 |
| Average | 0.7250 | 0.7250 | 0.7267 | 0.7503 |

TABLE VII
RECOGNITION ACCURACY WITH DIFFERENT TESTING DATA SUBSETS FROM THE EXTEND YALE B DATABASE.

| Methods | Right side shaded | Left side shade | Overall |
|---|---|---|---|
| RSC | 0.6491 | 0.6069 | 0.6280 |
| RRC_$L_1$ | 0.8202 | 0.7500 | 0.7851 |
| RRC_$L_2$ | 0.5887 | 0.6228 | 0.6058 |
| SSEC | 0.2237 | 0.2928 | 0.2583 |
| F-LR-IRNNLS | 0.7544 | 0.7325 | 0.7435 |
| VGGFace FC6 | 0.3874 | 0.3378 | 0.3626 |
| VGGFace FC7 | 0.3649 | 0.3198 | 0.3424 |
| Lightened CNN (A) | 0.4518 | 0.4211 | 0.4365 |
| Lightened CNN (B) | 0.5000 | 0.4912 | 0.4956 |
| Proposed GD-HASLR | 0.8553 | 0.7363 | **0.7960** |

From Table II, one can see that the proposed GD-HASLR algorithm has the best performance among ALL of the methods in ALL testing cases. It outperforms state-of-the-art methods for at least **5.08%** in the overall accuracy.

*B. Comparison with CNN-based Methods*

Moreover, in Table III, we compare the proposed algorithm to popular deep learning based methods. VGGFace-Net [24] is a VGG-Net [40] based deep convolutional neural network for face recognition. Together with the softmax method, it attains very high accuracy on face recognition. Lightened CNN [27, 28] is a newly released compact. They are effective CNN-based models for face recognition. It contains two models, A and B. Model A is inspired from the AlexNet [41] consisting of 4 max-pooling layers, 2 fully-connected (FC) layers, and 4 convolutional layers using the max feature map (MFM) as the activation function. Model B is inspired from the network in network (NIN) [42]. It consists of 4 max-pooling layers, 2 FC layers, 5 convolutional layers, and 4 NIN layers using the MFM as the activation function. The comparison of the proposed algorithm with these CNN based methods are shown in Table

III. Both the training datasets using one and two neutral faces for each person are examined.

From Table III, one can see that VGGFace outperforms most of state-of-the-art methods in Table II, including robust sparse coding-based methods. However, the proposed GD-HASLR algorithm still has the best performance in overall. We can see that the VGGFace has better performance on scarf occlusion but has bad performance on sunglasses occlusion. The analysis in [43][40] pointed out that the VGGNet is inefficient to upper face occlusions and here our experimental results accords with their conclusion. Also note that the performance of VGGFace does not become notably better when adding more training data. However, when using other methods, including the proposed one, from Tables I and II, more training data lead to better performance.

Similarly, when the lightened CNN is applied, the performance on the scarf subsets is better than that in the sunglasses subsets. Moreover, the model A performs better than the model B and also better than VGGFace. However, its performance is still not better than the proposed GD-HASLR algorithm in overall.

To illustrate the performance gap on different datasets of the CNN-based methods, one can observe the database images in the AR and the labeled face in the wild (LFW) [44] datasets. Only a small part of faces in LFW have occluded face data that contains some objects on the face. Within the occluded ones, little of them have over 20% occlusion rates. By contrast, in the AR dataset, the occlusion rates are approximately 25% for sunglasses, 40% for scarf. Therefore, although CNN-based algorithms have good performance in the LFW dataset, it does not mean that they are robust to face occlusion.

### C. Analysis

In Table IV, we justify that the adopted feature of generalized image gradient direction is effective and robust in the occluded face recognition problem. One can see that the performance of using the feature extracted in the image intensity domain is far worse than that of using the feature extracted from the generalized image gradient direction domain. Note that in the image intensity domain, occlusion can be characterized by a low-rank optimization problem, as explained in Section III-F. Nevertheless, occlusion in the generalized image gradient direction domain is a weak low-rank optimization problem. Then, from solving the weak low-rank optimization problem, the recognition performance is enhanced by leaps and bounds via robust gradient direction features.

From Table IV, results of using the generalized image gradient direction are far better than using image intensity. Both of them use the same regression model. This experiment shows that a better recognition performance can be achieved the weak low-rank optimization problem.

In Table IV, we also validate the proposed polling strategy using three different order image gradient directions as features. One can see that, the proposed polling method, which utilizes the discriminative power of different features to collaboratively decide the face identity, can further enhance the accuracy.

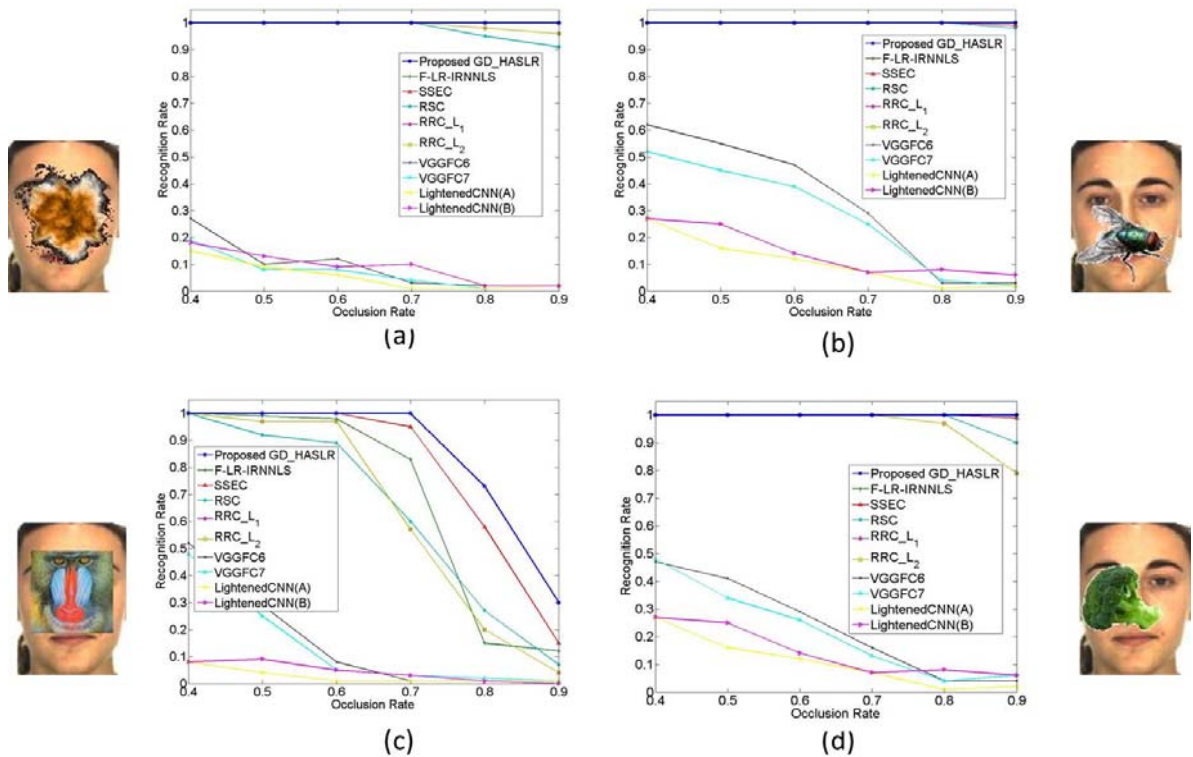

**Fig. 9.** Accuracy under different shapes and different sizes of occlusion patterns. (a) Sand or dust on the face. (b) A fly on the face with irregular shape. (c) A square baboon image on the face. (d) A broccoli on the face. The proposed GD-HASLR algorithm has the best performance in any case. In (c), the performances of other methods drop down quickly at the occlusion rate larger than 0.5 and the proposed GD-HASLR algorithm has much better performance even in the high occlusion rate case.

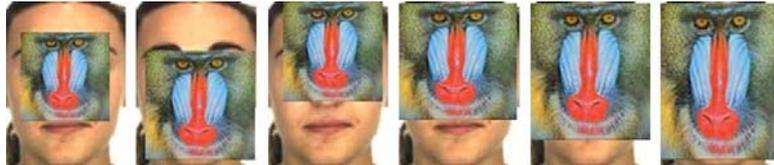

**Fig. 10.** Occlusion patterns with different occlusion rates. From left to right are 0.4, 0.5, 0.6, 0.7, 0.8, and 0.9, respectively.

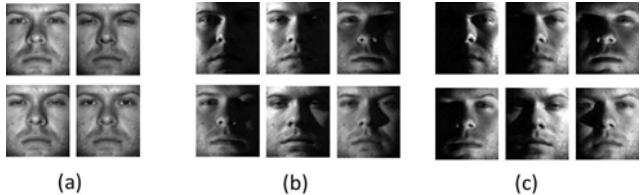

**Fig. 11.** Samples from the Extend Yale face dataset [45]. (a) Training dataset (neutral faces). (b) Left side shaded testing data. (c) Right side shaded testing data.

In Table V, different mapping functions, $\tan^{-1}(x)$, $\tanh(x)$, $\text{softsign}(x)$, and $\text{sigmoid}(x)$, are chosen to calculate the generalized image gradient features. Generally speaking, these four mapping functions have similar performances to project the data onto the image gradient direction domain. Simulations show that using $\tanh(x)$ can achieve the best performance, which is slightly over that of other three functions.

In Table VI, different feasible regularized weight functions associated with the hierarchical adaptive lasso in (17) are examined. The base method is with a constant weight [31], which leads to the original lasso problem. Much or less, using different functions is helpful for increasing the performance. The most suitable function is the normal inverse Gaussian (NIG), which can enhance the average accuracy by 2.503%.

### D. Synthetic Different Occlusion Patterns

Then, we conduct another experiment. We use synthesized occluded face data as testing data Different sizes, different shapes, and different locations of occlusions are used as occlusion patterns. We control the occlusion rate (i.e., the area of the occlusion pattern divided by the area of the face image) and the shape of occlusion to examine how occlusion patterns affect the recognition rate. Training data are neutral faces forming session 1 of the AR dataset. Four kinds of occlusion shapes are examined, which are shown in Fig. 9. The occlusion rates from 0.4 to 0.9 are tested in Fig. 10. Results are shown in Fig. 9. We compare to the proposed algorithm to FIRNNLS [22], RSC [4], RRC_$L_1$, RRC_$L_2$ [20], and deep learning-based methods, including VGGFace FC6, VGGFace FC7 [24], and lightened CNN models A and B [27, 28].

From Fig. 9, one can see that sparse coding based methods have high ability to resist outliers, even in the large occlusion rate case. In contrast, the deep learning based methods are not so robust to outliers. For the dense occlusion case, i.e., in the situation of high occlusion rates, the deep learning-based methods have limited performance. However, the proposed GD-HASLR algorithm still has very high recognition rates even if the face is seriously occluded. The proposed GD-HASLR algorithm has perfect results when the occlusion rate is smaller than 0.7. When the occlusion rate is larger than 0.7, GD-HASLR still has obvious better results than other methods. These experiment results validate that the proposed method is very robust to outliers even under heavy contiguous occlusion of arbitrary shape and location.

### E. Extend Yale Face Database for Shadow Occlusion

In addition to being robust to object occlusion, the proposed GD-HASLR algorithm is also robust to shadow.

In this subsection, we test the proposed method on illumination change caused shadow occlusion. Extend Yale face dataset [45] is the well-known face database associated with illumination change. This dataset contains 2432 face images of 38 people with 64 illumination variation conditions. We resize the image to 39x34 pixels. For each person, we select 4 nearly neutral face images as the training dataset. Moreover, 6 left side shaded face images and 6 right side shaded face images, which are obtained by the illumination from different angles, are selected as the testing dataset. Some training and testing samples are in Fig. 11. The results are shown in Table VII.

In Table VII, the proposed GD-HASLR algorithm has the best performance, which verifies that the proposed algorithm is also robust to shadow and the variation of illumination.

Since the proposed algorithm utilizes the gradient direction, which is invariant to pixel illumination change, the proposed algorithm can well recognize the face even if it is affected by shadow.

## V. CONCLUSION

In this paper, a very effective occluded face recognition algorithm, GD-HASLR, is proposed. It has strong robustness to the shape and the size of the occlusion object. It adopts generalized image gradient direction, which lets us have more choice on mapping functions and is a very robust feature for recognition. Next, we proposed the hierarchical sparse and low-rank regression model describing the sparsity on the coefficient vector and low-rankness on the error term. It also assigns adaptive weight on sparsity regularization. The model together with the image gradient direction feature lead to a weak low-rankness optimization problem, whose solution served as the reference map is weakly low-rank with blurred spatial structure. However, validated by experiments, the weak low-rankness optimization problem can have better recognition accuracy than in the image intensity domain, which is a normal low-rank optimization problem.

Simulation results show that the proposed GD-HASLR algorithm has strong robustness to any kind, any shape, and any size of occlusion, and have the best performance compared with state-of-the-art methods including robust sparse coding models and CNN-based methods.

## REFERENCES

[1] G. Hua, M. H. Yang, E. G. Learned-Miller, Y. Ma, M. Turk, D. J. Kriegman, and T. S. Huang, "Introduction to the special section on real-world face recognition," *IEEE Trans. Pattern Anal. Mach. Intell.*, vol. 33, no. 10, pp. 1921-1924, Oct. 2011.

[2] H. Jia and A. M. Martinez, "Support vector machines in face recognition with occlusions," *IEEE Conf. Computer Vision and Pattern Recognition*, pp. 136-141, 2009

[3] Z. Zhou, A. Wagner, J. Wright, H. Mobahi and Y. Ma, "Face recognition with contiguous occlusion using Markov random fields," *IEEE Int. Conf. Computer Vision*, pp.1050-1057, 2009.

[4] X. X. Li, D. Q. Dai, X. F. Zhang, and C. X. Ren, "Structured sparse error coding for face recognition with occlusion," *IEEE Trans. Image Processing*, vol. 22, no. 5, pp. 1889-1900, May 2013.

[5] R. Liang and X. X. Li, "Mixed error coding for face recognition with mixed occlusions," *Int. Joint Conf. Artificial Intelligence*, pp. 3657-3663, 2015.

[6] R. He, W. S. Zheng, T. Tan, and Z. Sun, "Half-quadratic based iterative minimization for robust sparse representation," *IEEE Trans. Pattern Anal. Mach. Intell.*, vol. 36, no. 2, pp. 261-275, Feb. 2014.

[7] J. Wright, A. Y. Yang, A. Ganesh, S. S. Sastry, and Y. Ma, "Robust face recognition via sparse representation," *IEEE Trans. Pattern Anal. Mach. Intell.*, vol. 31, no. 2, pp. 210-227, Feb. 2009.

[8] L. Zhang, M. Yang, and X. Feng, "Sparse representation or collaborative representation: Which helps face recognition?" *IEEE Int'l Conf. Computer Vision*, pp. 471-478, 2011.

[9] S. Cai, L. Zhang, W. Zuo, and X. Feng, "A probabilistic collaborative representation based approach for pattern classification," *IEEE Conf. on Computer Vision and Pattern Recognition*, 2016.

[10] M. Yang, L. Zhang, J. Yang, and D. Zhang, "Robust sparse coding for face recognition," *IEEE Conf. Computer Vision and Pattern Recognition*, pp. 625-632, 2011.

[11] M. Yang, L. Zhang, J. Yang, and D. Zhang, "Regularized robust coding for face recognition," *IEEE Trans. Image Processing*, vol. 22, no. 5, pp. 1753-1766, May 2013.

[12] M. Iliadis, H. Wang, R. Molina, and A. K. Katsaggelos, "Robust and low-rank representation for fast face identification with occlusions," arXiv preprint arXiv: 1605.02266, 2016.

[13] G. Liu, Z. Lin, S. Yan, J. Sun, Y. Yu and Y. Ma, "Robust recovery of subspace structures by low-rank representation," *IEEE Trans. Pattern Anal. Mach. Intell.*, vol. 35, no. 1, pp. 171-184, Jan. 2013.

[14] D. Zhang, Y. Hu, J. Ye, X. Li and X. He, "Matrix completion by truncated nuclear norm regularization," *IEEE Conf. Computer Vision and Pattern Recognition*, pp. 2192-2199, 2012.

[15] E. Candes and B. Recht, "Exact matrix completion via convex optimization," *Conf. Foundations of Computational Math.*, vol. 9, pp. 717-772, 2008.

[16] E. Elhamifar and R. Vidal, "Sparse subspace clustering: Algorithm theory and applications," *IEEE Trans. Pattern Anal. Mach. Intell.*, vol. 35, no. 11, pp. 2765-2781, Nov. 2013.

[17] G. Liu, H. Xu, and S. Yan, "Exact subspace segmentation and outlier detection by low-rank representation," *Int'l Conf. Artificial Intelligence and Statistics*, pp. 703-711, 2012.

[18] L. Ma, C. Wang, B. Xiao, and W. Zhou, "Sparse representation for face recognition based on discriminative low-rank dictionary learning," *IEEE Conf. Computer Vision and Pattern Recognition*, pp. 2586-2593, 2012.

[19] C. P. Wei, C. F. Chen, and Y. C. F. Wang, "Robust face recognition with structurally incoherent low-rank matrix decomposition," *IEEE Trans. Image Processing*, vol. 23, no. 8, pp. 3294-3307, Aug. 2014.

[20] J. Qian, L. Luo, J. Yang, F. Zhang, and Z. Lin, "Robust nuclear norm regularized regression for face recognition with occlusion," *Pattern Recognition*, vol. 48, no. 10, pp. 3145-3159, Oct. 2015.

[21] R. C. Gonzalez and R. E. Woods, *Digital Image Processing*, 3rd Edition, New Delhi: Pearson, 2008.

[22] A. Lee, F. Caron, A. Doucet, and C. Holmes, "A hierarchical Bayesian framework for constructing sparsity-inducing priors," *Technical Report*, University of Oxford, UK, pp. 1-18, 2010.

[23] A. Lee, F. Caron, A. Doucet, and C. Holmes, "Bayesian sparsity-path-analysis of genetic association signal using generalized *t* priors," *Statistical Applications in Genetics and Molecular Biology*, vol. 11, no. 2, pp. 1-29, 2012.

[24] Y. Taigman, M. Yang, M. A. Ranzato, and L. Wolf, "DeepFace: Closing the gap to human-level performance in face verification," *IEEE Conf. Computer Vision and Pattern Recognition*, pp. 1701-1708, 2014.

[25] F. Schroff, D. Kalenichenko, and J. Philbin, "Facenet: A unified embedding for face recognition and clustering," *IEEE Conf. Computer Vision and Pattern Recognition*, pp. 815-823, 2015.

[26] O. M. Parkhi, A. Vedaldi, and A. Zisserman, "Deep face recognition," *British Machine Vision Conference*, pp. 1–12, 2015.

[27] X. Wu, R. He, and Z. Sun, "A lightened CNN for deep face representation," arXiv preprint arXiv: 1511.02683, 2015.

[28] X. Wu, "Learning robust deep face representation," arXiv preprint arXiv: 1507.04844, 2015.

[29] D. G. Lowe, "Object recognition from local scale-invariant features," *Int'l Conf. Computer Vision*, pp. 1150-1157, 1999.

[30] G. Tzimiropoulos, S. Zafeiriou, and M. Pantic, "Sparse representations of image gradient orientations for visual recognition and tracking," *Proc. IEEE Conf. Computer Vision and Pattern Recognition Workshop*, pp. 26-33, 2012.

[31] K. P. Murphy, *Machine Learning: A Probabilistic Perspective*, MIT Press, Cambridge, MA, 2012.

[32] A. Y. Yang, S. S. Sastry, A. Ganesh, and Y. Ma, "Fast $L_1$-minimization algorithms and an application in robust face recognition: A review," *Int'l Conf. Image Processing*, pp. 1849-1852, 2010.

[33] E. Candes, X. Li, Y. Ma, and J. Wright, "Robust principal component analysis?," *Journal of the ACM*, vol. 58, issue 3, article 11, May 2011.

[34] S. Boyd, N. Parikh, E. Chu, B. Peleato, and J. Eckstein, "Distributed optimization and statistical learning via the alternating direction method of multipliers," *Foundations and Trends in Machine Learning*, vol. 3, no. 1, pp. 1-122, 2010.

[35] X. Xiang, M. Dao, Gregory D. Hager, and T. D. Tran, "Hierarchical sparse and collaborative low-rank representation for emotion recognition," *IEEE Int. Conf. Acoustics, Speech and Signal Processing*, pp. 3811-3815, 2015.

[36] J. Cai, E. Candes, and Z. Shen, "A singular value thresholding algorithm for matrix completion," *SIAM J. Optimization*, vol. 20, no. 4, pp. 1956-1982, 2010.

[37] D. L. Donoho, "Denoising by soft-thresholding," *IEEE Trans. Inf. Theory*, vol. 41, no. 3, pp. 613-627, Mar. 1995.

[38] O. E. Barndorff-Nielsen, "Normal inverse Gaussian distributions and stochastic volatility modeling," *Scand. J. Stat.*, vol. 24, pp. 1-13, Mar. 1997.

[39] A. M. Martinez, "PCA versus LDA," *IEEE Trans. Pattern Anal. Mach. Intell.*, vol. 23, no. 2, pp. 228-233, Feb. 2001.

[40] K. Simonyan and A. Zisserman, "Very deep convolutional networks for large-scale image recognition," *Proc. Int'l. Conf. Learn. Representations*, pp. 1-14, 2015.

[41] A. Krizhevsky, I. Sutskever, and G. E. Hinton, "ImageNet classification with deep convolutional neural networks," in *Advances in Neural Information Processing Systems*, pp. 1106–1114, 2012.

[42] M. Lin, Q. Chen, and S. Yan, "Network in network," *Computing Research Repository*, arXiv preprint arXiv: 1312.4400v3, 2013.

[43] M. M. Ghazi and H. K. Ekenel, "A Comprehensive analysis of deep learning based representation for face recognition," *IEEE Conf. Computer Vision and Pattern Recognition*, pp. 102–109, 2016.

[44] G. B. Huang, M. Ramesh, T. Berg, and E. Learned-Miller, "Labeled faces in the wild: A database for studying face recognition in unconstrained environments," *Technical Report 07-49 UMass*, vol. 1, pp. 1-11, 2007.

[45] A. Georghiades, P. Belhumeur, and D. Kriegman, "From few to many: illumination cone models for face recognition under variable lighting and pose," *IEEE Trans. Pattern Anal. Mach. Intell.*, vol. 23, issue 6, pp. 643–660, Jun. 2001.